\documentclass[12pt]{iopart}
\usepackage{graphicx}
\begin{document}
\newcommand{\be}{\begin{equation}}
\newcommand{\nn}{\nonumber}
\newcommand{\ee}{\end{equation}}
\newcommand{\bea}{\begin{eqnarray}}
\newcommand{\eea}{\end{eqnarray}}
\newcommand{\wee}[2]{\mbox{$\frac{#1}{#2}$}}   
\newcommand{\unit}[1]{\,\mbox{#1}}
\newcommand{\degree}{\mbox{$^{\circ}$}}
\newcommand{\ltish}{\raisebox{-0.4ex}{$\,\stackrel{<}{\scriptstyle\sim}$}}
\newcommand{\bin}[2]{\left(\begin{array}{c} #1 \\ #2\end{array}\right)}
\newcommand{\p}{_{\mbox{\small{p}}}}
\newcommand{\m}{_{\mbox{\small{m}}}}
\newcommand{\tra}{\mbox{Tr}}
\newcommand{\rs}[1]{_{\mbox{\tiny{#1}}}}        
\newcommand{\ru}[1]{^{\mbox{\small{#1}}}}

\title{Quantum probability rule: a generalisation of the theorems of Gleason and Busch}
\author{Stephen M. Barnett$^{1}$, James D. Cresser $^{2,3}$ John Jeffers$^{3}$ and David T. Pegg$^{4}$}

\address{$^{1}$School of Physics and Astronomy, University of Glasgow, Kelvin Building, University Avenue, Glasgow G12 8QQ, United Kingdom\\
$^{2}$Department of Physics and Astronomy, Faculty of Science,
Macquarie University, NSW 2109, Australia\\
$^{3}$Department of Physics, University of Strathclyde, Glasgow G4 0NG,
United Kingdom\\
$^{4}$Centre for Quantum Dynamics, Griffith University, Nathan, Brisbane, Queensland 4111, Australia}
\ead{john.jeffers@strath.ac.uk}

\begin{abstract}

Busch's theorem deriving the standard quantum probability rule can be regarded as a more general form of Gleason's theorem.  Here we show that a further generalisation is possible by reducing the number of quantum postulates used by Busch.  We do not assume that the positive measurement outcome operators are effects or that they form a probability operator measure. We derive a more general probability rule from which the standard rule can be obtained from the normal laws of probability when there is no measurement outcome information available, without the need for further quantum postulates.  Our general probability rule has prediction-retrodiction symmetry and we show how it may be applied in quantum communications and in retrodictive quantum theory.
\end{abstract}
\maketitle

\section{Introduction}
In the probabilistic interpretation of quantum measurement we have on one hand the physical process of preparing a system in some state and then performing a measurement procedure with the outcomes recorded, allowing probabilities which depend both on the measurement procedure and on the preparation process to be determined from the records of many experiments.    On the other hand we have the mathematics of Hilbert space entities.  To link the two it is axiomatic that there must be some postulate connecting a Hilbert space entity with something physical.  The standard quantum probability rule that does this has been highly successful for predicting the outcomes of measurements.  This rule could simply be accepted as the required postulate but it may be possible to obtain a better understanding of quantum theory if the rule could be deduced from more fundamental quantum postulates.  Gleason's theorem shows, given reasonable assumptions, that quantum probabilities must be expressible as expectation values of projectors or, more precisely, as the trace of the product of a projector and a density operator \cite{Gleason}.  This fundamental theorem is of central importance in quantum theory but although it is discussed in some textbooks \cite{Peres,Isham} a derivation of it rarely appears, doubtless because of the complexity of Gleason's proof.

Busch has provided a remarkable extension of Gleason's theorem \cite{Busch}.  It is remarkable in three ways: (i) it applies to state spaces of any dimension whereas Gleason's proof only applies for dimensions greater than two, (ii) it extends Gleason's proof by including generalised measurements \cite{Isham,Helstrom,Holevo,SMBBook} as well as projective ones and (iii) it is far simpler than Gleason's original proof \cite{Busch}.

Busch associates an outcome $m$ from a measurement with an effect $\hat{E}$, that is a positive operator less than the identity which can therefore be an element of a probability operator measure (POM), often also referred to as a positive operator-valued measure (POVM).  He equates the measurement outcome probability $p(m|s)$ for a system prepared in some state $s$ with the value of a function $v(\hat{E})$ which he requires to have the following three properties
\begin{eqnarray}
\nonumber
&({\rm P1})&  \qquad 0 \leq v(\hat{E}) \leq 1 \quad \forall \: \hat{E}    \\
\nonumber
&({\rm P2})&  \qquad v(\hat{\rm I}) = 1 \quad \hat{\rm I} = \;{\rm identity \; operator}  \\
\nonumber
&({\rm P3})& \qquad v(\hat{E}+\hat{F}+\cdots) = v(\hat{E}) + v(\hat{F}) + \cdots ,
\end{eqnarray}
where $\hat{F}, \cdots$ are also effects. The sum of the effects in (P3) must not exceed $\hat{\rm I}$. It should be emphasised that these properties are familiar in the theory of generalised measurements, but are derived on the basis of quantum theory \cite{Helstrom, Holevo}. Busch's aim was, and indeed ours is, rather different: the intention is to postulate these properties as \emph{axioms} and to \emph{derive}  quantum probabilities from them.  

It is not difficult to show from the normalisation condition for the probabilities of all possible outcomes combined with (P2) and the additivity condition (P3), that the sum of the effects must be the identity, that is, the unit operator. Thus the effects representing all possible outcomes for a system in state $s$ form a POM.  Also these conditions are consistent with the probability $p(m|s)$ given by $v(\hat{E})$  being non-contextual in the sense that it has this value independently of the particular POM to which $\hat{E}$ belongs, that is, it is independent of the particular measuring device as long as the outcome is represented by $\hat{E}$.  To see this, let $\hat{E}$ belong to two different POMs whose remaining elements are $\hat{F}_1, \hat{F}_2 \cdots$ and $\hat{G}_1, \hat{G}_2 \cdots$ corresponding to measuring devices $f$ and $g$ respectively. Then from normalisation and additivity we have
\bea
p(m|s,f) = 1-\sum_i v(\hat{F}_i) = 1- v\left( \sum_i \hat{F}_i \right)
\eea
with a corresponding expression for $p(m|s,g)$.  Because the elements of each POM must sum to the unit operator, $\sum_i \hat{F}_i = \sum_j \hat{G}_j$ so $p(m|s,f)=p(m|s,g)$.

Busch's property (P1) is a property of probabilities in general.  His quantum postulates, that is, those that concern Hilbert space operators, lie in (P2), (P3) and the association of a measurement outcome with an effect operator $\hat{E}$. In this paper we drop (P2) and weaken the effect quantum postulate so that it becomes a positive operator quantum postulate. This means we are {\it not} assuming that the operators representing the measurement outcomes are elements of a POM, which means that we no longer need to assume they are effects.  We do, however, assume they are bounded positive operators and adopt an additivity postulate similar to (P3) but we no longer limit the sum of measurement outcome operators to be $\leq \hat{\rm I}$.  We find that it is possible with this reduced number of quantum postulates to derive a probability rule that is more general than the standard rule $\tra (\hat{E}_i \hat{\rho})$. Furthermore we find that we can then deduce the standard rule from the general rule by the use of normal probability laws.

\section{General Probability Rule}
A measurement procedure for the determination of probabilities from a record of many experiments involving preparation and measurement will include a chosen measurement device and the method for recording the results obtained from it.  For example, two measurement events, such as a zero and a one photocount event, might be recorded as separate events or as a single event described as less than two photocounts. As another example, some experiments might not be recorded because of a post-selection procedure, whereby an experiment is ignored in the event of a particular measurement outcome.  We also include in the measurement procedure any means by which information can be obtained that affects the possibility of a recorded event.  This can include posterior knowledge.  For example if it is known that a photo-detector will be damaged if subjected to more than a certain number of photons, then an undamaged detector after the detection event will eliminate the possibility of a recording of a larger number of photons.  For our purposes here it is sufficient to specify a measurement procedure $x$ mathematically by the set of possible recorded measurement events $\{m_1, m_2, \cdots\}$ that can be obtained from it. We shall not be assuming non-contextuality with respect to the measurement procedure $x$ of the probability that a recorded measurement event is $m_i$, so we shall write this probability as $p(m_i|s,x)$ to show that it may depend on $x$ as well as the state $s$. 

Our first postulate is that, for a given measurement procedure $x$, each possible recorded event $m_i$ can be associated with a positive bounded Hilbert space \footnote{It suffices, for our purpose, to consider only state spaces of finite dimension. In this way we avoid complications such as observables with continuous spectra. We may incorporate such observables by means of a suitable limiting process, but such considerations would detract from the essential simplicity of the point that we are trying to make.} operator $\hat{M}_i$, in such a way that $p(m_i|s,x)$ is proportional to some function $u(\hat{M}_i)$ of this operator, that is, 
\bea
p(m_i|s,x) = Q(s,x) u(\hat{M}_i)
\eea
where the proportionality factor $Q(s,x)$ is the same for all $\hat{M}_i$ of the set of operators $\{\hat{M}_1, \hat{M}_2, \cdots \}$, which we can now use  to specify the measurement procedure $x$.  We are not assuming that $Q(s,x)$ is independent of the measurement procedure itself or of the particular state $s$.  We note that any set of positive bounded operators $\hat{M}_i$, to which we refer as measurement operators, can define mathematically a measurement procedure and that, while some measurement procedures have reasonably straightforward physical realizations, others may not.

The function $u(\hat{M}_i)$ may in general be a complex number $\exp(i\theta_i) w(\hat{M}_i)$, say, where $w(\hat{M}_i)$ is a positive number. The positivity of $p(m_i|s,x)$ for all $m_i$ then requires $Q(s,x)\exp(i\theta_i)$ to be positive for all $\theta_i$, which in turn requires $\theta_i$ all to have the same value which we write as $\theta$.  Thus we can, from (2), write our first postulate in the form
\bea
p(m_i|s,x) = N(s,x) w(\hat{M}_i)
\eea
where $N(s,x)$ is the positive normalisation factor $Q(s,x)\exp(i\theta)$. 

Our second postulate is that the positive function $w(\hat{A})$ of any positive bounded operator is additive, that is,
\bea
w(\hat{A}+\hat{B} + \cdots) = w(\hat{A}) + w(\hat{B}) + \cdots
\eea
for all positive bounded operators $\hat{A}, \hat{B}, \cdots$.  

We use a method similar to that used by Busch to show firstly that this additivity postulate implies linearity with respect to non-negative rational numbers.  From additivity we have, for positive integers $r$ and $n$, 
\bea
&&n w\left( \frac{r\hat{A}}{n}\right) = w (r\hat{A}) = rw(\hat{A}) \nonumber \\
\Rightarrow &&\frac{r}{n} w(\hat{A}) = w \left( \frac{r}{n} \hat{A} \right).
\eea
We can then use the additivity and positivity of $w(\hat{A})$ in a limiting argument similar to that used by Busch who showed that $\alpha v(\hat{E}) = v(\alpha \hat{E})$, where $\alpha$ is real and $0 \leq \alpha \leq 1$.  In our case we find that $\alpha w(\hat{A}) = w(\alpha \hat{A})$ where $\alpha$ is any non-negative real number. Combining this result with additivity we obtain the linearity relation
\bea
w\left( \sum_i \alpha_i \hat{M}_i \right) = \sum_i \alpha_i w(\hat{M}_i).
\label{addlin}
\eea

We are now in a position to prove our first main result.  The measurement operator $\hat{M}_i$ is a positive operator so we can write it in the diagonal form:
\begin{equation}
\hat{M}_i = \sum_\ell \lambda_\ell^i|\lambda_\ell^i\rangle \langle \lambda_\ell^i|,
\end{equation}
where $\{|\lambda_\ell^i\rangle\}$ are the eigenstates of $\hat{M}_i$ and $\lambda_\ell^i
= {\rm Tr}(\hat{M}_i |\lambda_\ell^i\rangle \langle \lambda_\ell^i|) \geq 0$ are the corresponding eigenvalues, which are all positive.  We should note that the positive operators $\{\hat{M}_i\}$ will, in general, be non-commuting and therefore will have distinct eigenvectors.  It follows, using our linearity condition (\ref{addlin}) that
\begin{equation}
w(\hat{M}_i) =  \sum_\ell {\rm Tr}(\hat{M}_i|\lambda_\ell^i\rangle \langle \lambda_\ell^i|) 
w(|\lambda_\ell^i\rangle \langle \lambda_\ell^i|).
\end{equation}
The $w(|\lambda_\ell^i\rangle \langle \lambda_\ell^i|)$  are simply positive numbers, however, and hence we can write 
\begin{equation}
w(\hat{M}_i) = \sum_\ell {\rm Tr}\left[\hat{M}_i |\lambda_\ell^i\rangle \langle \lambda_\ell^i| 
w(|\lambda_\ell^i\rangle \langle \lambda_\ell^i|)\right] = {\rm Tr}(\hat{M}_i \hat{R}_i),
\label{wmi1}
\end{equation}
where $\hat{R}_i$ is a positive operator, the diagonal elements of which, in the $\{ \lambda_\ell^i \}$ basis, are $w(|\lambda_\ell^i\rangle \langle \lambda_\ell^i|)$.  Equation (\ref{wmi1}) gives no such information about the off-diagonal elements of $\hat{R}_i$ so this operator is not completely determined by Eq. (\ref{wmi1}) but we can exploit the linearity relation (\ref{addlin}) to show that $\hat{R}_i$ must be independent of  $\hat{M}_i$ as follows.  Linearity and Eq. (\ref{wmi1}) require that $w(\hat{M}_1)+ w(\hat{M}_2)$ equals $\tra[ (\hat{M}_1 +\hat{M}_2) \hat{R}_{12} ]$ so we must be able to write $w(\hat{M}_1)$ and $w(\hat{M}_2)$ in the form $\tra (\hat{M}_1 \hat{R}_{12})$ and $\tra(\hat{M}_2 \hat{R}_{12})$ respectively, where the common operator $\hat{R}_{12}$ has diagonal elements $w(|\lambda_\ell^1\rangle \langle \lambda_\ell^1|)$ in the $\{ \lambda_\ell^1 \}$ basis and $w(|\lambda_\ell^2\rangle \langle \lambda_\ell^2|)$ in the $\{ \lambda_\ell^2 \}$ basis.  We can combine $\hat{M}_1$ with {\it any} other positive bounded operators to form a set defining a measurement procedure so the common operator $\hat{R}$ must have diagonal elements $w(|\lambda_\ell\rangle \langle \lambda_\ell|)$ in {\it any} basis $\{\lambda_\ell \}$ and thus is independent of any particular $\hat{M}_i$.  We can then write
\bea
w(\hat{M}_i) = \tra(\hat{M}_i \hat{R})
\eea
for all $\hat{M}_i$, showing that the probability that a measurement event is $m_i$ depends both on the associated measurement operator and an independent operator, which it is natural to associate physically with the preparation process.  We can show that common operator $\hat{R}$ is unique by using the lemma that two operators having the same diagonal elements in all bases must be equal. We prove this lemma in the Appendix.

To obtain the probability $p(m_i|s,x)$ we require the proportionality factor $N(s,x)$, which can be found from the normalisation condition that the probabilities of all possible outcomes sum to unity.  This yields 
\bea
p(m_i|s,x) = \frac{\tra(\hat{M}_i \hat{R})}{\tra(\hat{X} \hat{R})}
\eea
where $\hat{X} = \sum_j M_j$.  Dividing the numerator and denominator by $\tra(\hat{R})$ yields our general probability law
\bea
p(m_i|s,x) = \frac{\tra(\hat{M}_i \hat{\rho})}{\tra(\hat{X} \hat{\rho})}.
\label{genproblaw}
\eea
We note that $\hat{X}$ depends only on the possible recorded measurement outcomes, thereby characterising the particular measurement procedure $x$, leaving the unit-trace positive $\hat{\rho}$ as a density operator to characterise the state $s$.  This is the first main result of the paper: if we reduce the number of Busch's quantum postulates by discarding (P2) and, relaxing the assumption that the operator representing a measurement outcome must be an effect to simply being a positive bounded operator, we arrive at a probability law that any set of positive operators (with finite eigenvalues) can provide a set of probabilities and that these probabilities are calculated using (\ref{genproblaw}).

Before proceeding, we give a simple illustration of the meaning of our second postulate, the additivity postulate.  The measurement procedure $x$ only enters into (\ref{genproblaw}) as the sum $\hat{X}$. Consider a particular measuring device with, among other measurement events $m_3, m_4, \cdots$, the events $m_1$ and $m_2$ corresponding to $\hat{M}_1$ and $\hat{M}_2$ if these are recorded separately.  If we record these events together as one event $m_1$ {\it or} $m_2$ our additivity postulate implies that the corresponding measurement operator is $\hat{M}_1+\hat{M}_2$. The sum $\hat{X}$ is thus the same whether the measurement procedure involves separately recorded events or a single combined event. As a result of this, while $p(m_3|s,x)$ depends on whether $m_4$ is a possible recorded event or not, it does not depend on whether $m_1$ and $m_2$ are recorded together or separately. 

\section{Standard probability rule}
It remains for us to determine the physical meaning of our general probability law. In doing so we arrive, very naturally, at a Bayesian interpretation. Consider the case where we know that a number of possible states $s_k$, for which the density operators are $\hat{\rho}_k$, have probabilities $p_k$ of being the prepared state. The state $s$ based on this knowledge will have a density operator $\hat{\rho} = \sum_k p_k \hat{\rho}_k$ representing the average or {\it a priori} density operator and the probability of the recorded event being $m_i$ will be given by (\ref{genproblaw}).  If the state actually prepared was $s_k$, say, then in place of  (\ref{genproblaw}) we would have a different probability 
\bea
p(m_i|s_k,x) = \frac{\tra(\hat{M}_i \hat{\rho}_k)}{\tra(\hat{X} \hat{\rho}_k)}.
\label{problaw1}
\eea
We should be able to obtain (\ref{genproblaw}) as a sum of these objects, suitably weighted by a probability:
\bea
p(m_i|s,x) &=& \sum_k p(m_i|s_k,x) P_k \nonumber \\
\Rightarrow \tra(\hat{M}_i \hat{\rho}) &=&  \sum_k \tra(\hat{M}_i \hat{\rho}_k) P_k \frac{\tra(\hat{X} \hat{\rho})}{\tra(\hat{X} \hat{\rho}_k)}\nonumber \\
&=& \sum_k \tra(\hat{M}_i \hat{\rho}_k) p_k.
\eea
For this to hold in general we need only to set
\bea
P_k = \frac{\tra(\hat{X} \hat{\rho}_k)}{\tra(\hat{X} \hat{\rho})} p_k.
\label{pkreln}
\eea
The fact that both the $P_k$ and the  $p_k$ are probabilities means that their ratio is a likelihood \cite{Box}, which we can interpret as the likelihood of $s_k$  given $x$:
\bea
l(s_k|x) = \frac{\tra(\hat{X} \hat{\rho}_k)}{\tra(\hat{X} \hat{\rho})}.
\eea
In order to adopt this interpretation it is necessary to interpret $P_k$ as an {\it a posteriori} probability based on some knowledge relating to the recorded outcome of the measurement.  Specifically this will be knowledge affecting the possibility that some outcomes may occur.  As $\hat{X}$ is the sum of the operators representing the possible recorded measurement outcomes, its value will depend on this knowledge.  Then $P_k$ will also depend on this knowledge from Eq. (\ref{pkreln}).  The simplest example is where the actual outcome itself is known to be $m_i$, say, and then $\hat{X}=\hat{M}_i$ as no other outcomes are possible any longer.  We then find from Eq. (\ref{genproblaw}) that the {\it a posteriori} probability that the outcome is $m_i$ is unity as it must be.  Another example is where joint events $(s, m)$ showing the input state and the consequent measurement outcome are recorded after a known post-selection procedure has rejected some joint events containing particular measurement outcomes.  This has the effect of reducing the number of possible recorded outcomes and thus the sum of the operators representing them.  In this context $P_k$ is just the probability that the state in a recorded joint event is $s_k$.  

We shall express the {\it a posteriori} nature of $P_k$ by writing it as $P(s_k|x)$, that is, the probability that state $s_k$ was prepared in a recorded experiment conditioned on the operator corresponding to the measurement outcome being limited to one of the reduced number of terms in the posterior expression for $\hat{X}$.  This leads us in turn to interpret $p(m_i|s,x)$ in Eq. (\ref{genproblaw}) as 
\bea
p(m_i|s,x) &=& \frac{\tra(\hat{M}_i \hat{\rho})}{\tra(\hat{X} \hat{\rho})} = \sum_k \frac{\tra(\hat{M}_i \hat{\rho}_k)}{\tra(\hat{X} \hat{\rho}_k)} P(s_k|x) \nonumber \\
&=& \sum_k p(m_i|s_k,x) P(s_k|x),
\eea
which is consistent with Bayesian probability, confirming our interpretation of $P(s_k|x)$. 

If there is no post-selection and no posterior knowledge about measurement results that can eliminate or reduce the possibility of particular measurement events and thus of the preparation events that may have produced them, then the {\it a posteriori} probability $P(s_k|x)$  that any state $s_k$ has occurred must be equal to the {\it a priori} probability $p_k$ that this state occurs.  In this case we have, from Eq. (\ref{pkreln})
\bea
\tra( \hat{X} \hat{\rho}_k) = \tra( \hat{X} \hat{\rho})
\label{post=pri}
\eea
for all $\hat{\rho}_k$.  Consider two density operators $\hat{\rho}_k$ with $k = 1, 2$ related by a unitary transformation $\hat{\rho}_2 = \hat{U} \hat{\rho}_1 \hat{U}^{-1}$.  From Eq. (\ref{post=pri}) we then have
\bea
\tra (\hat{U} \hat{X} \hat{\rho}_1 \hat{U}^{-1}) = \tra(\hat{X} \hat{\rho}_1) = \tra (\hat{X} \hat{\rho}_2) = \tra (\hat{X} \hat{U} \hat{\rho}_1 \hat{U}^{-1} ).
\eea
For this to hold for any $\hat{\rho}_1$, $\hat{X}$ must commute with any $\hat{U}$ and must therefore be proportional to the unit operator, that is $\hat{X}=K \hat{\rm I}$.  Then our general probability rule (\ref{genproblaw}) becomes the standard, or restricted, probability law
\bea
p(m_i|s) = \tra ( K^{-1} \hat{M}_i \hat{\rho} ) = \tra ( \hat{E}_i \hat{\rho} )
\label{buschrelate1}
\eea
say, where
\bea
\sum_i \hat{E}_i = K^{-1} \sum_i \hat{M}_i = K^{-1} \hat{X}  = \hat{\rm I}.
\eea
$\hat{E}_i$  are therefore effects and form a POM.  In Busch's notation $p(m_i|s)$ equals $v(\hat{E}_i)$. Using the latter expression for the left side of Eq. (\ref{buschrelate1}) and then summing both sides over $i$ gives condition (P2), which we see is a {\it result} of our approach, obtained from our general formula (\ref{genproblaw}) by the usual rules of probability,  rather than being an additional quantum postulate. 

\section{Applications}
It is natural to ask whether there are any applications of our more general probability formula (\ref{genproblaw}).  Here we present three such applications.  An obvious, but often overlooked, one is to measurement probabilities when we have some (incomplete) information about the measurement outcome.  It is often the case in quantum optics experiments, for example, that we restrict our attention to probabilities given some future event, such as a two-photon cascade in which the detection of one photon is used to herald the emission of another \cite{Grangier}.  In such cases $\hat{X}$ will be restricted to only those measurement event operators $\hat{M}_i$ that include the heralding event.

A second example arises in the theory of quantum communications \cite{SMBBook}.  Here a transmitting party, Alice, selects from a set of possible states $s_i$, with density operators $\hat{\rho}_i$ and prior selection probabilities $p_i$, and sends a quantum system prepared in this state to a receiving party, Bob.  Bob's task is to determine from a measurement, as well as possible, the state prepared by Alice.  As he knows from the measurement that the outcome is $m_j$ corresponding to  $\hat{M}_j$, say, he knows that $\hat{X}$ contains just this single term, that is, his knowledge has eliminated the possibility of all other terms.  He can therefore simply write the sum of the possible terms as $\hat{X} = \hat{M}_j$ and obtain from (\ref{pkreln}) the {\it a posteriori}, or retrodictive, probability that Alice sent the system in state $s_k$
\bea
P(s_k|m_j) = \frac{ \tra (\hat{M}_j \hat{\rho}_k p_k) }{\tra (\hat{M}_j \hat{\rho} )}.
\label{retlaw}
\eea
We note that retrodictive probabilities such as this can also be found by using Bayes' theorem in conjunction with the usual expression for the quantum probability $\tra (\hat{E}_j \hat{\rho})$ \cite{BPJRetr}.  Peres \cite{Peres2} has described an expression equivalent to Eq. (22) as the only retrodictive form that can be legitimately derived from conventional quantum mechanics.  However here there is no need to add a Bayes rule; it is already contained in the general probability law (\ref{genproblaw}) expressed in the form (\ref{pkreln}).  We note that there is symmetry between the retrodictive form of our probability law (\ref{retlaw}) and the predictive form (\ref{genproblaw}) which we write here as
\bea
p(m_k|s_j,x) = \frac{\tra (\hat{\rho}_j \hat{M}_k)} {\tra(\hat{\rho}_j \hat{X})}
\label{predform}
\eea
with $\hat{\rho}_k p_k$ in (\ref{retlaw}) corresponding to $\hat{M}_k$ in (\ref{predform}), $\hat{M}_j$ in (\ref{retlaw}) corresponding to $\hat{\rho}_j$ in (\ref{predform}) and thus $\hat{\rho}$ in (\ref{retlaw}) corresponding to $\hat{X} = \sum_k \hat{M}_k$ in (\ref{predform}).  This allows Bob an alternative and equivalent way to retrodict by defining a density operator $\hat{\rho}_j$ for a ``retrodictive state" as $\hat{M}_j/\tra(\hat{M}_j)$, writing $\hat{M}_k$ as $\hat{\rho}_k p_k$ and writing $\hat{X}$ as $\hat{\rho}$ and then substituting into the right side of  the predictive formula (\ref{predform}) to obtain the retrodictive expression (\ref{retlaw}).  In this way the general probability rule (\ref{genproblaw}) can be used for {\it both} prediction and retrodiction without the need to invoke Bayes' theorem, which is already effectively contained in the law.  If there is a time interval between preparation and measurement, then Alice would need to allow for evolution of her predictive state in this interval to calculate the probability of a measurement event and Bob would need to allow for the retroevolution of the retrodictive state to retrodict a preparation event.

Our final example completes the resolution of a long-standing controversy in retrodictive quantum theory \cite{Retrodiction}. In retrodictive quantum theory we assign a retroevolving quantum state on the basis of a {\it later} measurement and can use this to ask questions about, among other things, initial preparation events. It has been suggested that we can only apply quantum retrodiction if there is no prior information about the preparation event so the prior initial density operator has an unbiased form and is proportional to the identity operator \cite{Belinfante, Amri}. This is a result of attempting to find a retrodictive formula by making the restricted predictive probability $\tra(\hat{E}_i \hat{\rho})$ symmetric or causally neutral \cite{leiferspekkens} or by using a time-reversed form of Gleason's theorem \cite{Amri}. From the symmetry inherent in our general probability rule, which reduces to the restricted predictive form when $\hat{X} \propto \hat{ \rm I}$, it is easy to see from the correspondence between $\hat{X}$ and $\hat{\rho}$ above that our general retrodictive formula will reduce to the restricted retrodictive form when  $\hat{\rho} \propto \hat{\rm I}$. To obtain the general, and far more useful, retrodictive probability formula from causal neutrality of a predictive formula it is necessary to start with the general predictive form. For this reason it is also inadequate to use a time-reversed form of Gleason's theorem. Looked at from another view point, retrodicted preparation probabilities are quite often contextual, depending on what other states could possibly be prepared. For example if photon number states are being prepared, there is some limit set by the amount of energy available or simply by the difficulty in preparing some states\footnote{This type of situation is considered by Dressel and Jordan \cite{Dressel}, who use a symmetric formulation of quantum theory based on quantum instruments (basically corresponding to measurement devices) to derive predictive, retrodictive and ``interdictive" states.}. Thus time-reversed theorems incorporating non-contextuality are inappropriate for a general treatment. 

\section{Conclusion}
We should note that it is also possible to derive a relationship between Bayes' theorem, predictive and retrodictive quantum theory based on an {\it assumed} expression for measurement and preparation probabilities in which preparation and measurement operators appear symmetrically \cite{PrepMeas,opensys}.  Our general probability rule as derived in this paper, however, enables us to arrive at the correct expression for retrodictive probabilities {\it without} postulating a symmetric form for the probabilities and thus may be regarded as a more fundamental approach that formally justifies this earlier work.

Busch has relaxed Gleason's postulate that measurement outcomes must be represented by projectors by allowing measurement outcomes to be represented by effects.  In this paper we have further relaxed this to the postulate that the probability of a measurement outcome for a particular input state and measurement procedure is proportional to a positive additive function of a bounded positive operator.  By allowing the proportionality constant to depend not just on the state but also on the measurement procedure, including choice of measurement device, we are explicitly not assuming non-contextuality in relation to other possible measurement outcomes.  Any set of positive operators (with strictly finite eigenvalues) can represent measurement outcomes and can be used to calculate the probabilities of these outcomes.  The usually adopted requirement that these operators must sum to the identity is not assumed but {\it follows} from our approach for the case when there is no prior information about the measurement outcome.  This resulting standard, or restricted, probability formula is seen to be a special case of a more general causally-neutral  symmetric formula, which can be used for both prediction, that is finding probabilities of measurement outcomes, and for retrodiction involving finding the probabilities of preparation events.  When used for prediction, the formula is applicable even when there is partial knowledge of possible measurement outcomes as may occur when post-selection is involved or when there is simply incomplete reporting of outcomes that have occurred.  Retrodictive probabilities can, of course, be calculated from the usual restricted formula by employing Bayes' theorem but there is no need to invoke Bayes' theorem when using the general formula.  Important examples of the use of our general formula include quantum communications, retrodictive quantum theory and where there is prior agreed postselection of measurement results.

\section*{Acknowledgements}
This work was supported by the Royal Society, Wolfson Foundation and the UK EPSRC.  DTP thanks D. Matthew Pegg for his support.\\

\section*{Appendix: Uniqueness Lemma}

Here we seek to prove that two operators with the same diagonal elements in all bases must be equal. 

Assume that we have two operators $\hat{R}$ and $\hat{Q}$ with the same diagonal elements in any basis. If this is true then for any pair of basis states $|i\rangle$ and $|j\rangle$ we have 
\bea
\langle i| \hat{R} |i \rangle &=& \langle i| \hat{Q} |i \rangle, \label{ii}\\
\langle j| \hat{R} |j \rangle &=& \langle j| \hat{Q} |j \rangle. \label{jj}
\eea
If the diagonal elements are the same in any basis then, for a general superposition of these states
\bea
|u\rangle = a|i\rangle + b|j\rangle,
\eea
with $a$ and $b$ being any pair of complex amplitudes, we must now also have
\bea
\langle u| \hat{R} |u \rangle = \langle u| \hat{Q} |u \rangle.
\eea
If this is true for all $a$ and $b$ then it must also be true for the coefficients of $ab^*$ and $a^*b$ in this expression so that
\bea
\langle i| \hat{R} |j \rangle &=& \langle i| \hat{Q} |j \rangle, \label{ij}\\
\langle j| \hat{R} |i \rangle &=& \langle j| \hat{Q} |i \rangle. \label{ji}
\eea
There is nothing special about the states we have chosen and hence we infer that all the matrix elements of the operators are equal in this basis.  If all the matrix elements are equal then the operators must be identical.
\end{document}